# S-band acoustoelectric amplifier utilizing an ultra-high thermal conductivity heterostructure for low self-heating

L. Hackett,[1] X. Du,[2] M. Miller,[1] B. Smith,[1] S. Santillan,[1] J. Montoya,[1] R. Reyna,[1] S. Arterburn,[1] S. Weatherrend,[1] T. A. Friedmann,[1] R. H. Olsson III,[2a)] and M. Eichenfield[1,3a)]

[1]*Microsystems Engineering, Science, and Applications, Sandia National Laboratories, Albuquerque, NM, 87123, USA*
[2]*Department of Electrical and Systems Engineering, University of Pennsylvania, Philadelphia, PA, 19104, USA*
[3]*College of Optical Sciences, University of Arizona, Tucson, AZ, 85719, USA*

Here we report on an acoustoelectric slab waveguide heterostructure for phonon amplification using a thin $Al_{0.58}Sc_{0.42}N$ film grown directly on a 4H-SiC substrate with an ultra-thin $In_{0.53}Ga_{0.47}As$ epitaxial film heterogeneously integrated onto the surface of the $Al_{0.58}Sc_{0.42}N$. The aluminum scandium nitride film grown directly on silicon carbide enables a thin (1 micron thick) piezoelectric film to be deposited on a thermally conductive bulk substrate (370 W/m-K for 4H-SiC), enabling negligible self-heating when combined with the $In_{0.53}Ga_{0.47}As$ semiconductor parameters of large mobility (~7000 cm$^2$/V-s) and low concentration of charge carriers (~5x10$^{15}$ cm$^{-3}$). A Sezawa mode with optimal overlap between the peak of its evanescent electric field and the semiconductor charge carriers is supported. The high velocity of the heterostructure materials allows us to operate the Sezawa mode amplifier at 3.05 GHz, demonstrating a gain of 500 dB/cm (40 dB in 800 μm). Additionally, a terminal end-to-end radio frequency gain of 7.7 dB and a nonreciprocal transmission of 52.6 dB are achieved with a dissipated DC power of 2.3 mW. The power added efficiency and acoustic noise figure are also characterized.

---

a) Electronic mail: rolsson@seas.upenn.edu, eichenfield@arizona.edu

The development of technologically relevant phononic amplifiers utilizing the acoustoelectric effect, which occurs when the electric field of a piezoelectric acoustic wave interacts with a semiconductor charge carrier system, has been ongoing since initial demonstrations in the 1960s.[1,2,3] The main impediment has been achieving large gain with low self-heating. In this type of device, a voltage is applied to the charge carriers inducing a drift current, and the kinetic energy in the drift carriers can be transferred to the acoustic wave through a time-delayed Coulombic interaction. However, the drift current also generates Ohmic power in the semiconductor whose primary heat dissipation pathway is conduction through the substrate. The resultant self-heating has only recently been mitigated to operate acoustoelectric amplifiers with terminal gain in continuous operation.[3-5] This has been achieved with two strategies: 1) reduction of carrier concentration to reduce the field and power required to reach a given gain and 2) increasing substrate thermal conductivity to dissipate heat more effectively. The first strategy limits the saturation output power directly, which is critical to these devices operating in realistic radio frequency systems; so efforts to achieve low self-heating by further improving the substrate thermal conductivity are desirable.

Along with high thermal conductivity, the substrate must support integration with strongly piezoelectric thin films, have a high acoustic velocity for acoustic confinement in the piezoelectric, and have a high resistivity for low radio frequency loss. The electromechanical coupling coefficient, $k^2$, of the acoustic mode is a parameter that quantifies the fraction of total energy stored in the electric field,[6] which partly determines both the gain slope (gain per volt) and completely determines the maximum achievable gain in an acoustoelectric phonon amplifier in the absence of self-heating and other gain clamping mechanisms. To achieve a high $k^2$ while maintaining low heterostructure thermal resistivity, a strongly piezoelectric thin film must be integrated with sufficient overlap between the evanescent electric field of the acoustic mode and the semiconductor charge carriers.

Here we report the characterization of a phononic amplifier at a center frequency of 3.05 GHz that combines both strategies, lower carrier concentration and higher thermal conductivity substrate, to achieve low self-heating. This is accomplished in a material stack of an indium gallium arsenide ($In_{0.53}Ga_{0.47}As$) epitaxial layer bonded to a substrate of 1 μm thick alloyed aluminum scandium nitride ($Al_{0.58}Sc_{0.42}N$) film grown on 4H silicon carbide (4H-SiC). An equivalent circuit model[6] can be utilized to express the dissipated DC power required to achieve peak gain in an acoustoelectric phononic amplifier as a function of the semiconductor parameters. According to this theory, the electric field required to achieve peak gain, $E_{max}$, is given by

$$E_{max} = \frac{v_a}{\mu} + \frac{Nqd}{(\varepsilon_p + \varepsilon_0)} \qquad (1)$$

where $v_a$ is the acoustic velocity, $\mu$ is the charge carrier mobility, $N$ is the charge carrier concentration, $q$ is the elementary charge, $d$ is the semiconductor thickness, $\varepsilon_p$ is the piezoelectric permittivity, and $\varepsilon_0$ is the vacuum permittivity. The first term in Eqn. (1), $v_a/\mu$, is the field required for the charge carriers to drift at a velocity equal to the acoustic velocity. At this point, no acoustic loss or amplification is achieved. Acoustic gain occurs at fields exceeding $v_a/\mu$ and theoretically increases up to the field given in Eqn. (1), although experimentally various mechanisms, such as heating effects,[2,4] prohibit achieving peak gain. Utilizing Eqn. (1), we can determine an expression for the dissipated DC power to achieve maximum gain, $P_{DC,max}$,

$$P_{DC,max} = \frac{qN\mu wd}{L}\left(\frac{v_a L}{\mu} + \frac{NqdL}{(\varepsilon_p + \varepsilon_0)}\right)^2 \quad (2)$$

where $w$ and $L$ are the semiconductor width and length, respectively. From Eqn. (2), minimizing the charge carrier concentration, $N$, significantly reduces the dissipated power. The dependence on the mobility, $\mu$, is more complicated in that increasing $\mu$ reduces the required field to achieve gain, but also increases the generated drift current at a given voltage. Therefore, it is typically advantageous for $\mu$ to be above a certain threshold value. For example, in the devices studied in this work, a mobility of >5400 cm$^2$/V-s is required to achieve gain with an applied bias of 10 V or less. Larger voltages both increase the DC power requirements and limit applicability of devices. Beyond the threshold to lower the applied voltages, it is advantageous to increase $\mu$ only if the drift current can be minimized through other means.

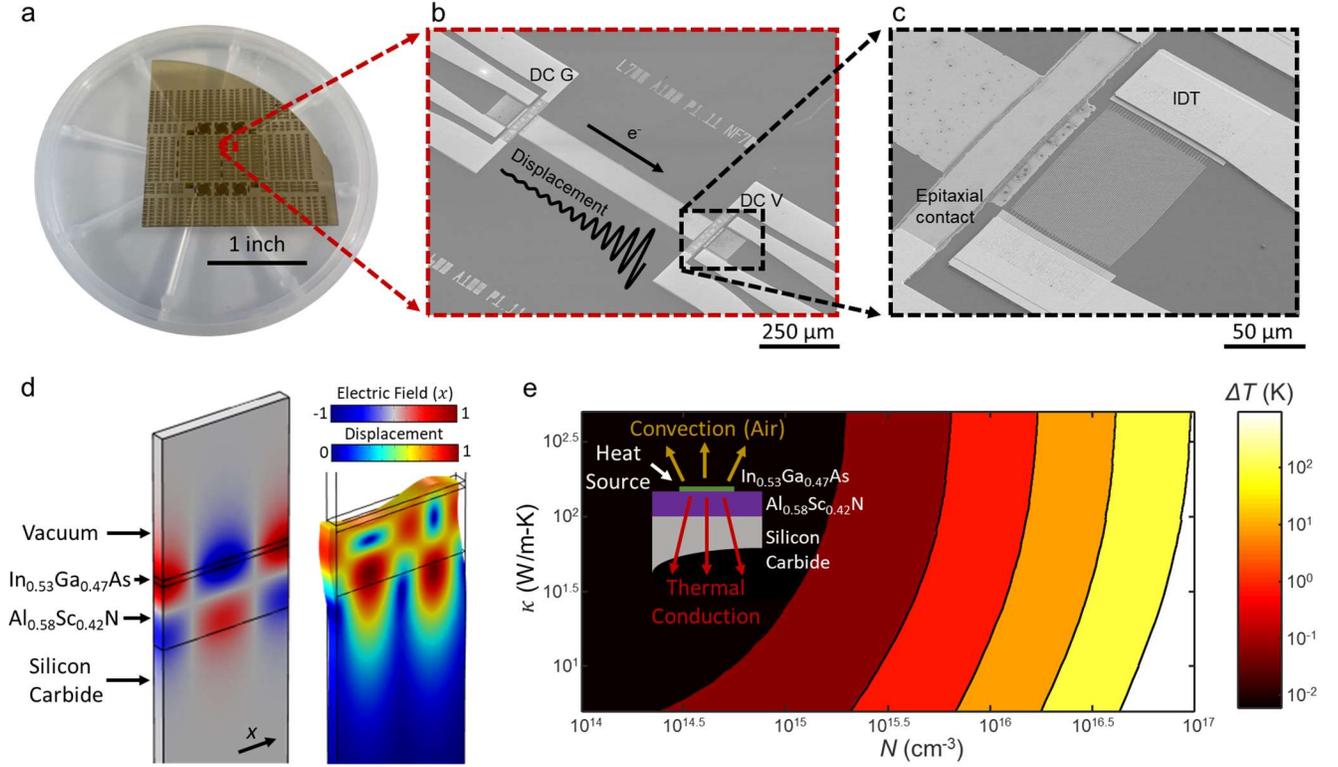

FIG. 1. (a) Quarter wafer of phonon amplifiers and delay lines fabricated in an $In_{0.53}Ga_{0.47}As/Al_{0.58}Sc_{0.42}N/SiC$ acoustoelectric heterostructure. (b) SEM image of a single amplifier device. The doped $In_{0.53}Ga_{0.47}As$ semiconductor layer, which is in contact with the $Al_{0.58}Sc_{0.42}N$ surface and patterned between the input and output IDTs, serves as the source of charge carriers. An external applied bias causes the electrons to drift at a velocity that exceeds the acoustic wave velocity, which leads to phononic amplification via the interaction between the electrons and the evanescent electric field of the piezoelectric acoustic wave. (c) SEM image showing additional details of the fabricated IDT and DC contact structure. (d) Finite element method models showing the Sezawa mode's longitudinal electrical field and displacement in the full material stack. (f) Plot of the temperature increase ($\Delta T$) to achieved 40 dB of acoustic gain as a function of semiconductor charge carrier concentration, $N$, and substrate thermal conductivity, $\kappa$. The inset shows a schematic of heat generation and dissipation in the system.

The $In_{0.53}Ga_{0.47}As$ epitaxial semiconductor utilized in this work provides a large charge carrier mobility (~7000 cm$^2$/V-s) and a low concentration of charge carriers (~5x10$^{15}$ cm$^{-3}$), such that large gain can be achieved with low dissipated power. A strongly guided acoustic mode (Sezawa mode)[7] with an acoustic velocity of ~6700 m/s in the $In_{0.53}Ga_{0.47}As/Al_{0.58}Sc_{0.42}N/SiC$ stack is supported due to the high acoustic velocity in SiC. The 4H-SiC substrate also provides a large thermal conductivity (370 W/m-K),[8] which reduces the heterostructure thermal resistance to enable efficient heat transfer. With this developed material platform, we demonstrate 40 dB of acoustic gain (7.7 dB of terminal end-to-end radio frequency gain) while dissipating 2.3 mW of DC power. The amplifier is highly nonreciprocal, with ≥52.6 dB of nonreciprocal transmission demonstrated at this operating point. Additional characterization includes the DC to radio frequency power conversion efficiency and acoustic noise figure.

A camera image of a quarter wafer of fabricated $In_{0.53}Ga_{0.47}As/Al_{0.58}Sc_{0.42}N/SiC$ phononic amplifiers, $Al_{0.58}Sc_{0.42}N/SiC$ acoustic delay lines, and Hall structures is shown in Fig. 1(a). Fabrication begins by growing an In$_{0.53}$Ga$_{0.47}$As/InP lattice-matched heterostructure via metal organic chemical vapor deposition[4] on an InP substrate. Separately, the Al$_{0.58}$Sc$_{0.42}$N film is grown on a SiC substrate, as reported previously,[7] followed by bonding between the In$_{0.53}$Ga$_{0.47}$As/InP and Al$_{0.58}$Sc$_{0.42}$N/SiC via manual initiation and overnight annealing at 350°C. The InP substrate is then removed via wet etching followed by forming the epitaxial contacts, amplifier layer, DC metal contacts, and IDTs as discussed in previous work.[4,9] A single phononic amplifier is shown in the scanning electron microscope image in Fig. 1(b) with additional details of the epitaxial contact and IDT shown in Fig. 1(c). The longitudinal electric field and the displacement field for the Sezawa mode in the In$_{0.53}$Ga$_{0.47}$As/Al$_{0.58}$Sc$_{0.42}$N/SiC are shown in Fig. 1(d) and Fig. 1(e), respectively. As can be seen, the peak of the longitudinal electric field, which is the field primarily responsible for the acoustoelectric interaction, overlaps with the In$_{0.53}$Ga$_{0.47}$As, leading to optimal electron-phonon coupling.

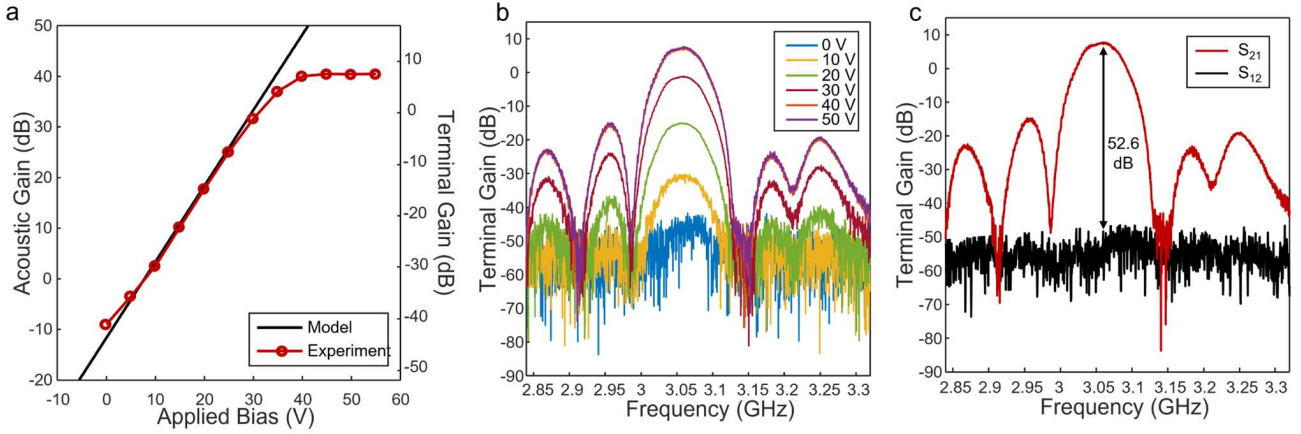

FIG. 2. (a) Acoustic and terminal gain as a function of applied bias for an acoustic wave amplifier with an interaction length of 800 μm. A maximum acoustic gain of 40 dB and a terminal (end-to-end radio frequency) gain of 7.7 dB are achieved at an applied bias of 45 V. (b) Measured terminal gain as a function of frequency for different applied bias values. (c) Transmission for the forward propagating ($S_{21}$) and backward propagating ($S_{12}$) acoustic waves at an applied bias of 45 V. A nonreciprocal transmission of 52.6 dB is achieved at this operating point.

Of existing models in the literature, the acoustoelectric interaction in this material stack is most accurately described by a perturbative approach[10] which gives the following expression for the acoustoelectric gain coefficient, $\alpha_m$

$$\alpha_m = \frac{1}{2} \frac{(v_d/v_a - 1)\omega_c \varepsilon_s w Z_a \beta \tanh(\beta d)}{(v_d/v_a - 1)^2 + (R\omega_c/\omega + H)^2} \quad (3)$$

where $Z_a$ is the interaction impedance, $w$ is the interaction width, $\omega_c$ is the dielectric relaxation frequency, $\varepsilon_0$ is the vacuum permittivity, $\varepsilon_s$ is the semiconductor permittivity, $d$ is the semiconductor thickness, $R$ is the space-charge reduction factor, and $H$ is a term that arises from semiconductor charge carrier diffusion.[10] By combining a finite element method thermal model with Eqn. (3), we can characterize the steady-state temperature rise, $\Delta T$, to achieve acoustic gain. Fig. 1(e) shows a plot of $\Delta T$ at 40 dB of acoustic gain as a function of the concentration of charge carriers, $N$, and the substrate thermal conductivity, $\kappa$. The inset of Fig. 1(e) shows a schematic of heat generation and removal in the heterostructure. As can be seen, the values of $N$ and $\kappa$ modify $\Delta T$ over several orders of magnitude and therefore minimization of deleterious self-heating effects can be achieved through optimization of these two parameters alone.

Figure 2(a) shows a plot of the measured acoustic and terminal gain as a function of applied bias. All measurements were made with a network analyzer on a custom probe station after a short-load-open-through calibration. The modeled acoustic gain (Eqn. 3) is also shown. For the model, all parameters except for $k^2$, which modifies the interaction impedance, $Z_a$, were taken from the experiment or known values from the literature. With this approach, we found that a $k^2$ value of $(7 \pm 3)$% best fits the experimental data where the uncertainty in $k^2$ results from the uncertainty in the other parameters utilized in Eqn. 3. At 45 V, a maximum acoustic gain of 40 dB is achieved (terminal end-to-end radio frequency gain of 7.7 dB) at a DC power dissipation of 2.3 mW. While experimentally the acoustic gain saturates beyond a drift voltage of 45 V, this is not captured in the modeled gain curve. We also observe clamping of the current at the same bias values where gain clamping occurs. A potential mechanism for the gain saturation is that, at a drift velocity of ~4x10$^6$ cm/s, we have reached the saturation velocity for the charge carriers in this acoustoelectric heterostructure, such that increasing the bias does not further increase the ratio between the drift and acoustic velocities and therefore the gain remains fixed at a constant value. Figure 2(b) shows the measured terminal gain as a function of frequency for different applied bias values. The fractional bandwidth is 1.6% at a bias of 50 V.

The measured $S_{21}$ and $S_{12}$ as a function of frequency are show in Fig. 2(c) with an applied bias of 45 V. As can be seen, a nonreciprocal transmission contrast of ≥52.6 dB is achieved between the acoustic wave that propagates in the same direction as the electrons, and is therefore amplified, and the backwards propagating acoustic wave, which is attenuated. The value of 52.6 dB is taken directly from the measured scattering parameters on a network analyzer. As the network analyzer settings are not optimized to capture large attenuation, it is possible for the actual nonreciprocal transmission to be >52.6 dB.

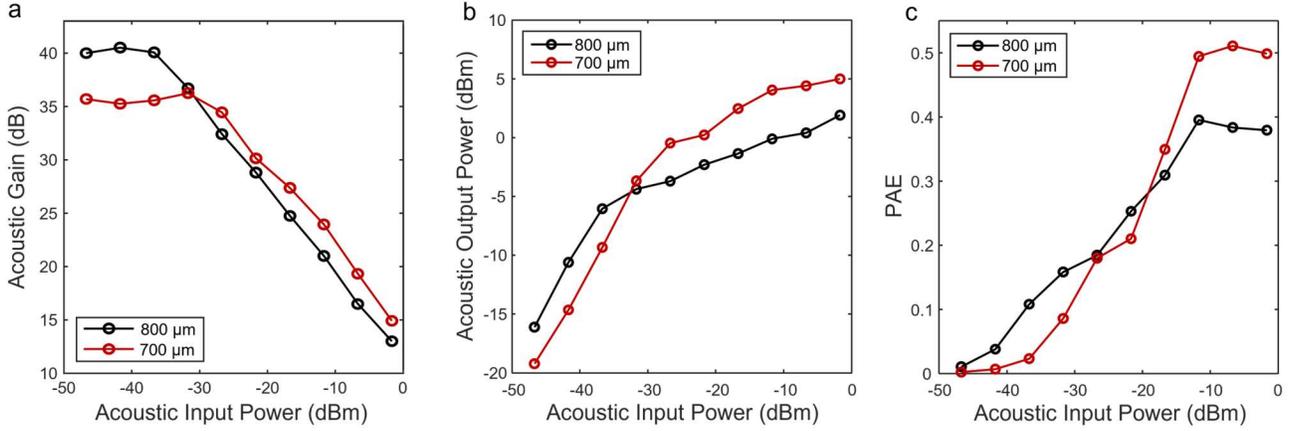

FIG. 3. (a) Acoustic gain as a function of acoustic input power for devices with acoustoelectric interaction lengths of 700 μm and 800 μm. (b) Acoustic output power as a function of acoustic input power and (c) power added efficiency (PAE) as a function of acoustic input power for the same two devices.

The data shown in Fig. 2 was taken with an acoustic input power of -41.7 dBm and therefore we have shown that the device provides large phononic gain with low DC power dissipation for low input powers. Next, we examine the performance of the device as the input acoustic power is varied. Figure 3(a) shows a plot of the acoustic gain as a function of the acoustic input power while Fig. 3(b) shows a plot of the acoustic output power as a function of the acoustic input power for amplifiers with acoustoelectric interaction lengths of 700 μm and 800 μm. For the 700 μm length, there is an onset of gain compression at an acoustic input power of -27.6 dBm (acoustic output power of 0.5 dBm) while for the 800 μm length, gain compression occurs at an acoustic input power of -31.7 dBm (acoustic output power of -4.6 dBm). A plot of the power added efficiency as a function of acoustic input power for the same two interaction lengths is shown in Fig. 3(c). The power added efficiency, PAE, is defined as $PAE = \frac{P_{A,OUT} - P_{A,IN}}{P_{DC,IN}}$ where $P_{A,OUT}$ is the acoustic output power, $P_{A,IN}$ is the acoustic input power, and $P_{DC,IN}$ is the DC input power. Here we demonstrate a power added efficiency of 10% at an acoustic gain of 40 dB. To maximize the power added efficiency, the objective is to achieve large gain at large acoustic power with low DC power. As can be seen in Fig. 3(c), we do achieve relatively large power added efficiency values (>50%) but in regions of high gain compression. As the fundamental mechanism for gain saturation in these types of devices is the polarization of all available charge carriers,[10] likely the charge carrier concentration will need to be increased and other mechanisms for achieving large gain and low DC power dissipation must be explored. This could include reducing the semiconductor width (see Eqn. (2)) or increasing the gain slope through operating at a higher frequency.

Another critical metric for an acoustoelectric phonon amplifier is the noise figure, which quantifies the degradation to the signal-to-noise ratio.[11] Figure 4 shows a plot of the acoustic noise figure as a function of acoustic gain. The acoustic noise figure reports the noise figure for the isolated acoustoelectric interaction region alone. We find that at an acoustic gain of 40 dB, the acoustic noise figure is $10 \pm 1$ dB for the device with the acoustoelectric interaction length of 800 μm. For the device with an acoustoelectric interaction length of 700 μm, the acoustic noise figure is $7 \pm 1$ dB at an acoustic gain of 35 dB. The noise figure for both devices is significantly larger than the ~1.5 dB acoustic noise figure predicted by theory.[12] Additional research is required to identify and reduce contributions to the noise figure for the acoustoelectric heterostructure approach in this material stack and will be the focus of future work.

In conclusion, we have reported on the design, fabrication,

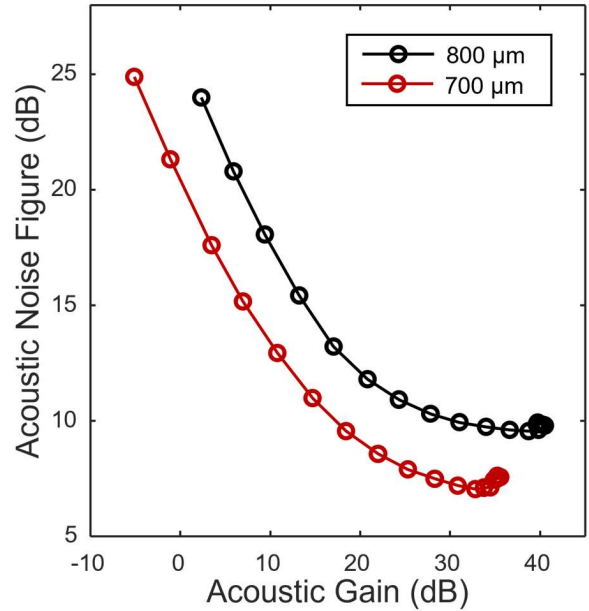

FIG. 4. A plot of acoustic noise figure as a function of acoustic gain for phononic amplifier devices with 700 μm and 800 μm acoustoelectric interaction lengths.

and characterization of phononic amplifiers with a center frequency of 3.05 GHz and low self-heating in a material stack of In$_{0.53}$Ga$_{0.47}$As/Al$_{0.58}$Sc$_{0.42}$N/SiC. We have demonstrated an acoustic gain of 500 dB/cm (40 dB for an 800 μm acoustoelectric interaction length) and a terminal radio frequency gain of 7.7 dB with a dissipated DC power of 2.3 mW. The growth of alloyed Al$_{0.58}$Sc$_{0.42}$N films on a SiC substrate is promising for the development of active acoustic wave devices as it provides the ability to achieve thin (1 μm or less) piezoelectric films directly on a thermally conductive substrate with a high resistivity and large acoustic velocity. The combination of material parameters supports both high frequency (>1 GHz) operation and low self-heating. Two critical areas of future study for phononic amplifiers that operate based on the acoustoelectric effect include the exploration of gain compression and saturation output power in addition to a more detailed study of the noise figure to close the gap between the experimental and theoretical values.

This research was developed in part with funding from the Defense Advanced Research Projects Agency (DARPA). This work was carried out in part at the Singh Center for Nanotechnology at the University of Pennsylvania, a member of the National Nanotechnology Coordinated Infrastructure (NNCI) network, which is supported by the National Science Foundation (Grant No. HR0011-21-9-0004). The authors performed this work, in part, at the Center for Integrated Nanotechnologies, an Office of Science User Facility operated for the U.S. Department of Energy (DOE) Office of Science. Support was provided by the Laboratory Directed Research and Development program at Sandia National Laboratories, a multimission laboratory managed and operated by National Technology and Engineering Solutions of Sandia, LLC., a wholly owned subsidiary of Honeywell International, Inc., for the U.S. Department of Energy's National Nuclear Security Administration under Contract No. DE-NA-003525. This paper describes objective technical results and analysis. Any subjective views or opinions that might be expressed in the paper do not necessarily represent the views of the U.S. Department of Energy or the United States Government.

## DATA AVAILABILITY
The data that support the findings of this study are available from the corresponding author upon reasonable request.

### AUTHOR DECLARATIONS
**Conflict of Interest**
The authors have no conflicts to disclose.

### Author Contributions
**Lisa Hackett:** Conceptualization (lead); Investigation (lead); Methodology (lead); Writing – original draft (lead); Writing – review & editing (equal). **Xingyu Du:** Investigation (supporting). **Michael Miller:** Investigation (supporting). **Brandon Smith:** Investigation (supporting). **Steven Santillan:** Investigation (supporting); Methodology (supporting). **Josh Montoya:** Investigation (supporting); Methodology (supporting). **Robert Reyna:** Investigation (supporting). **Shawn Arterburn:** Investigation (supporting). **Scott Weatherrend:** Investigation (supporting). **Tom Friedmann:** Investigation (supporting); Supervision (supporting). **Roy Olsson:** Conceptualization (lead); Supervision (lead) **Matt Eichenfield:** Conceptualization (lead); Supervision (lead); Writing – review & editing (equal); Funding acquisition (lead)

# REFERENCES

1. Christopher P Carmichael, M Scott Smith, Arthur R Weeks, and Donald C Malocha, *IEEE transactions on ultrasonics, ferroelectrics, and frequency control* **65** (11), 2205 (2018); LA Coldren and GS Kino, Applied Physics Letters **23** (3), 117 (1973); Larry Allen Coldren and GS Kino, *Applied Physics Letters* **18** (8), 317 (1971); Siddhartha Ghosh, *Journal of Micromechanics and Microengineering* **32** (11), 114001 (2022); L Hackett, A Siddiqui, D Dominguez, JK Douglas, A Tauke-Pedretti, T Friedmann, G Peake, S Arterburn, and M Eichenfield, *Applied Physics Letters* **114** (25) (2019); H Hanebrekke and KA Ingebrigtsen, *Electronics Letters* **16** (6), 520 (1970); Hakhamanesh Mansoorzare and Reza Abdolvand, presented at the 2021 IEEE 34th International Conference on Micro Electro Mechanical Systems (MEMS), 2021; RK Route and GS Kino, *IBM Journal of Research and Development* **13** (5), 507 (1969).
2. Lisa Hackett, Michael Miller, Felicia Brimigion, Daniel Dominguez, Greg Peake, Anna Tauke-Pedretti, Shawn Arterburn, Thomas A Friedmann, and Matt Eichenfield, *Nature communications* **12** (1), 2769 (2021).
3. Donald C Malocha, Christopher Carmichael, and Arthur Weeks, *IEEE Transactions on Ultrasonics, Ferroelectrics, and Frequency Control* **67** (9), 1960 (2020).
4. Lisa Hackett, Michael Miller, Scott Weatherred, Shawn Arterburn, Matthew J Storey, Greg Peake, Daniel Dominguez, Patrick S Finnegan, Thomas A Friedmann, and Matt Eichenfield, *Nature Electronics* **6** (1), 76 (2023).
5. Hakhamanesh Mansoorzare and Reza Abdolvand, *IEEE Transactions on Microwave Theory and Techniques* **70** (11), 5195 (2022).
6. Robert Adler, *IEEE Transactions on Sonics and Ultrasonics* **18** (3), 115 (1971).
7. Xingyu Du, Zichen Tang, Chloe Leblanc, Deep Jariwala, and Roy H Olsson, presented at the 2022 Joint Conference of the European Frequency and Time Forum and IEEE International Frequency Control Symposium (EFTF/IFCS), 2022.
8. Michael E Levinshtein, Sergey L Rumyantsev, and Michael S Shur, *Properties of Advanced Semiconductor Materials: GaN, AlN, InN, BN, SiC, SiGe*. (John Wiley & Sons, 2001).
9. Lisa Hackett, Matthew Koppa, Brandon Smith, Michael Miller, Steven Santillan, Scott Weatherred, Shawn Arterburn, Thomas A Friedmann, Nils Otterstrom, and Matt Eichenfield, arXiv preprint arXiv:2305.01600 (2023).
10. GS Kino and TM Reeder, *IEEE Transactions on Electron Devices* **18** (10), 909 (1971).
11. Harald T Friis, *Proceedings of the IRE* **32** (7), 419 (1944).
12. GS Kino and LA Coldren, *Applied Physics Letters* **22** (1), 50 (1973).